\documentclass[twocolumn]{aastex631}
\usepackage{mathtools}
\begin{document}

\title{Photospheric velocity evolution of SN 2020bvc: signature of $r$-process nucleosynthesis from a collapsar}

\author[0000-0002-8391-5980]{Long Li}
\affiliation{Deep Space Exploration Laboratory / Department of Astronomy, University of Science and Technology of China, Hefei 230026, China; lilong1125@ustc.edu.cn, daizg@ustc.edu.cn}
\affiliation{School of Astronomy and Space Science, University of Science and Technology of China, Hefei 230026, China}

\author[0000-0002-1766-6947]{Shu-Qing Zhong}
\affiliation{Deep Space Exploration Laboratory / Department of Astronomy, University of Science and Technology of China, Hefei 230026, China; lilong1125@ustc.edu.cn, daizg@ustc.edu.cn}
\affiliation{School of Astronomy and Space Science, University of Science and Technology of China, Hefei 230026, China}

\author[0000-0002-7835-8585]{Zi-Gao Dai}
\affiliation{Deep Space Exploration Laboratory / Department of Astronomy, University of Science and Technology of China, Hefei 230026, China; lilong1125@ustc.edu.cn, daizg@ustc.edu.cn}
\affiliation{School of Astronomy and Space Science, University of Science and Technology of China, Hefei 230026, China}
\affiliation{School of Astronomy and Space Science, Nanjing University, Nanjing 210023, China}

\begin{abstract}
Whether binary neutron star mergers are the only astrophysical site of rapid neutron-capture process ($r$-process) nucleosynthesis remains unknown. Collapsars associated with long gamma-ray bursts (GRBs) and hypernovae are promising candidates. Simulations have shown that outflows from collapsar accretion disks can produce enough $r$-process materials to explain the abundances in the universe. However, there is no observational evidence to confirm this result at present. SN 2020bvc is a broad-lined type Ic (Ic-BL) supernova (SN) possibly associated with a low-luminosity GRB. Based on semi-analytic SN emission models with and without $r$-process materials, we perform a fitting to the multi-band light curves and photospheric velocities of SN 2020bvc. We find that in a $r$-process-enriched model the mixing of $r$-process materials slows down the photospheric recession and therefore matches the velocity evolution better. The fitting results show that $r$-process materials with mass of $\approx0.36~M_\odot$ and opacity of $\approx4~\rm cm^2~g^{-1}$ is needed to mix with about half of the SN ejecta. Our fitting results are weakly dependent on the nebular emission. Future statistical analysis of a sample of type Ic-BL SNe helps us understand the contribution of collapsars to the $r$-process abundance.

\end{abstract}


\section{Introduction} \label{sec:intro}

The astrophysical origin of rapid neutron-capture process ($r$-process) nucleosynthesis responsible for the production of about half of the elements heavier than iron remains an open question (e.g., \citealt{Cowan1991,Arnould2007,Qian2007,Thielemann2011,Cowan2021,Siegel2022}). Many studies (see \citealt{Cowan2021}, for a review) suggest that the extremely high neutron densities ($\gtrsim 10^{26}~\rm{cm^{-3}}$) required for $r$-process nucleosynthesis can be realized in the core collapse of massive stars or in compact binary mergers involving at least one neutron star.

GW170817, the first gravitational-wave event from a binary neutron star merger, was detected by the gravitational-wave observatories Advanced Laser Interferometer Gravitational-Wave Observatory and Advanced Virgo Interferometer on August 17, 2017 \citep{Abbott2017a}. Subsequently, a low-luminosity short gamma-ray burst (GRB) GRB 170817A was detected about 1.7 seconds after the coalescence \citep{Abbott2017b,Goldstein2017,Savchenko2017}. Roughly 11 hours later, an luminous optical counterpart AT2017gfo was discovered in the galaxy NGC 4993 \citep{Coulter2017}. Dozens of ground and space telescopes followed up the optical counterpart in the ultraviolet, optical, and near-infrared bands for about 30 days after the merger \citep{Arcavi2017,Chornock2017,Coulter2017,Cowperthwaite2017,Drout2017,Evans2017,Kasliwal2017,Nicholl2017,Pian2017,Smartt2017,Soares-Santos2017,Tanvir2017}. The luminosity and the multiband light-curve evolution support the hypothesis that AT2017gfo is a kilonova powered by the radioactive decay of $r$-process nuclei synthesized within the neutron-rich merger ejecta \citep{Li1998,Metzger2010,Roberts2011,Barnes2013,Rosswog2018}. Furthermore, \cite{Watson2019} directly identified the neutron-capture element strontium in the spectra of AT2017gfo. Observations of GW170817, GRB 170817A, and AT2017gfo provide the first direct indication that $r$-process elements can be produced in binary neutron star merger.

On the other hand, Galactic chemical evolution models have been challenging neutron star mergers as the only, if not the dominant, sites of $r$-process nucleosynthesis (e.g., \citealt{Wehmeyer2015,Cote2019,Haynes2019,vandeVoort2020,Molero2021}). The rate of neutron star mergers seems too low and the delay times relative to star formation too long to explain the $r$-process elemental abundances of the extremely metal-poor stars in the Milky Way. The astrophysical sites associated with core-collapse supernovae (SNe) have long been promising candidates as the source of main $r$-process contributions. For regular core-collapse SNe, previous theoretical work has mainly focused on $r$-process nucleosynthesis in neutrino-driven winds from proto-neutron stars (e.g., \citealt{Takahashi1994,Woosley1994}). However, both theoretical studies (unlikely high entropy and/or moderate electron fraction environments; e.g., \citealt{Qian1996,Thompson2001,Roberts2012,Martinez-Pinedo2012,Curtis2019}) and observations (incompatible with production yields and occurrence frequencies; e.g., \citealt{Sneden2008,Roederer2014,Hansen2018}) suggest that regular core-collapse SNe are unlikely to undergo strong $r$-processes producing heavy elements up to the actinides, and at most weak $r$-processes synthesising elements below europium. Recent studies have linked the $r$-process to two rare types of SNe: magnetorotational/jet-driven SNe (e.g., \citealt{Nishimura2015,Nishimura2017}), and SNe from collapsars (e.g., \citealt{Siegel2019,Barnes2022}).

Collapsars, the collapse of rapidly rotating massive stars, are expected to form black hole-accretion disk systems that produce long GRBs and hypernovae \citep{MacFadyen1999}. \cite{Siegel2019} found that at sufficiently high accretion rates, the matter in the mid-plane of the collapsar accretion disk is driven into a neutron-rich state due to charged-current weak interactions, thus leading to a rapidly expanding neutron-rich disk outflows sufficient to undergo strong $r$-process nucleosynthesis. However, when adopting different configurations (e.g. the treatment of neutrino absorption) for numerical simulations, the conditions of low electron fraction required for strong $r$-process are not always be expected 
\citep{Miller2020,Fujibayashi2022,Just2022a,Just2022b}. Due to the high opacity of the $r$-process material, it is possible to find observational signatures of the $r$-process in SN light curves and spectra. Depending on the degree of mixing of the outflows with the ejecta, excess near-infrared emission is expected to emerge at different stages of the SN light curves \citep{Siegel2019,Barnes2022}. Recently, \cite{Anand2023} performed a systematic study of 25 broad-lined type Ic (Ic-BL) SNe (which are most likely to be associated with collapasars) with optical/near-infrared bands. However, fitting results show no convincing evidence of $r$-process enrichment in their sample.

The SN photospheric velocity evolution derived from the spectra allows independent constraints on the properties of ejecta and explosions. Previous studies have shown that $^{56}$Ni mixing can affect the photospheric velocity evolution in the stripped-envelope SNe (e.g., \citealt{Dessart2012,Moriya2020}). In general, with increased $^{56}$Ni mixing, the photosphere recedes slower, thus leading to the photosphere velocity staying at a higher level over time. It can be expected that, due to the high opacity of the $r$-process material, the mixing of $r$-process material may lead to a similar effect on the photospheric velocity.

SN 2020bvc is an Ic-BL SN discovered on February 4, 2020 by the All Sky Automated Survey of Supernovae (ASAS-SN) in the nearby ($z=0.025235$) galaxy UGC 9379 \citep{Perley2020,Stanek2020}. In addition to optical observations, radio and X-ray counterparts have been detected by Very Large Array, Swift-XRT and Chandra, respectively \citep{Izzo2020,Ho2020a}. The results of the search for historical GRB data show that there is no typical GRB associated with 2020bvc, but the presence of a low luminosity GRB is still possible \citep{Ho2020b}. Considering the very high expansion velocity and the X-ray light curve, \cite{Izzo2020} suggests that 2020bvc is associated with an off-axis GRB. However, the radio observations seem to be inconsistent with the off-axis GRB scenario \citep{Ho2020b}. \cite{Ho2020b} also present a double-peaked feature in the optical light curve derived from the Zwicky Transient Facility data, where the first peak can be attributed to the shock cooling emission. The most noticeable feature of SN 2020bvc is its spectral  evolution. Although the early spectra of SN 2020bvc is similar to that of other SNe Ic-BL (e.g., SN 1998bw), however, the overall shape of the spectra unexpectedly remains almost constant from about day 20 to day 73, which is quite different from other SNe Ic-BL \citep{Ho2020b,Rho2021}. The measured Fe II $\lambda$5169 absorption velocities between day 20 and day 73 are almost identical, which are about $20,000~\rm{km~s^{-1}}$ \citep{Ho2020b,Rho2021}.

Recently, \cite{Barnes2022} studied the signatures of $r$-process enrichment in the light curves of SNe from collapsars based on a semi-analytic model. In this paper, we focus on the features of $r$-process enrichment in the photospheric velocity evolution. Based on semi-analytic model, we fit the multi-band light curves and photospheric velocities of SN 2020bvc. In Section \ref{sec:model}, we introduce our semi-analytic model. In Section \ref{sec:results}, we fit the light curves and velocity evolution of SN 2020bvc using the semi-analytic model and the Monte Carlo Markov chain approach, and present the fitting results. Our discussion and conclusions can be found in Section \ref{sec:discussion}.

\section{The model} \label{sec:model}

The basic framework of our $r$-process-enriched SN emission model is similar to \cite{Barnes2022}. After the SN explosion, a homologously expanding spherical ejecta is assumed. The radioactive $^{56}$Ni synthesized in the explosion and the $r$-process-rich outflows from the collapsar accretion disk are mixed with the ejecta to a certain degree due to various hydrodynamic processes. The observed emission consists of photospheric emission from the photosphere and nebular emission from the optically thin region. Under the assumption of homologous expansion, the photospheric velocity can be easily derived from the photospheric radius. Next we show the details of our model.

\subsection{Density Profile} \label{subsec:density}

A few seconds after the SN explosion,  the ejecta begins to enter the homologous expansion phase after expanding to a few times its radius. The density profile of the SN ejecta can be approximated as a segmented power law, with a steep power law at large radii and a flatter power law at small radii. Given the total kinetic energy ($E_{\rm SN}$) and total mass ($M_{\rm ej}$) of the ejecta, the density profile can be expressed as \citep{Chevalier1982,Chevalier1989,Matzner1999,Kasen2010}
\begin{equation}
\label{eq:density}
\rho(v, t)=
\begin{dcases}
\zeta_\rho \frac{M_{\mathrm{ej}}}{v_{\mathrm{t}}^3 t^3}\left(\frac{r}{v_{\mathrm{t}} t}\right)^{-\delta} & v<v_{\mathrm{t}}, \\ 
\zeta_\rho \frac{M_{\mathrm{ej}}}{v_{\mathrm{t}}^3 t^3}\left(\frac{r}{v_{\mathrm{t}} t}\right)^{-n} & v>v_{\mathrm{t}},
\end{dcases}
\end{equation}
where $v_t$ is the transition velocity which can be derived from the density continuity condition,
\begin{equation}
v_{\mathrm{t}}=\zeta_v\left(\frac{E_{\mathrm{SN}}}{M_{\mathrm{ej}}}\right)^{1 / 2}.
\end{equation}
The coefficients $\zeta_\rho$ and $\zeta_v$ are given by
\begin{equation}
\zeta_\rho=\frac{(n-3)(3-\delta)}{4 \pi(n-\delta)},
\end{equation}
\begin{equation}
\zeta_v=\left[\frac{2(5-\delta)(n-5)}{(n-3)(3-\delta)}\right]^{1 / 2}.
\end{equation}
A dimensionless radius of the break in the SN ejecta density profile from the inner flatter component to the outer steeper component $x_0$ is introduced to link $v_t$ and the characteristic velocity ($v_{\rm SN}$) of the SN ejecta \citep{Chatzopoulos2012},
\begin{equation}
v_{\mathrm{SN}}=\frac{v_{\mathrm{t}}}{x_0}.
\end{equation}

Equation (\ref{eq:density}) is valid only if $n>5$ and $\delta<3$. The parameter $n$ depends on the properties of the progenitor and the convection in the outer layers of the star. For the progenitors of type Ib/c and Ia SNe, one has $n\simeq10$ \citep{Matzner1999,Kasen2010,Moriya2013}. For the progenitors of red supergiants, one has $n\simeq12$ \citep{Matzner1999,Moriya2013}. Typical values of the slope of the inner density profile of the ejecta is generally taken as $\delta\simeq0-2$. In our model, we adopt $\delta=1,n=10$ as default values and obtain $\zeta_\rho=0.124,\zeta_v=1.69$.

\subsection{Material Mixing} \label{subsec:mixing}

Due to various hydrodynamic instabilities (e.g., Rayleigh-Taylor instability), large-scale mixing may occur between the core and the envelope of the ejecta, the collapsar disk outflows and the ejecta. The degree of mixing is highly uncertain. In particular, mixing of the radioactive $^{56}$Ni and $r$-process material alters the internal energy and opacity profiles, thus changing the observational behaviour of SNe. In our model, we put $^{56}$Ni and $r$-process material in the ejecta with different degrees of mixing. We introduce a maximum dimensionless mixing radius of $^{56}$Ni ($x_{\rm Ni}$) and $r$-process material ($x_{\rm rp}$). Under $x_{\rm Ni}$ and $x_{\rm rp}$, $^{56}$Ni and $r$-process material are evenly distributed within the ejecta. Considering a more realistic scenario, we impose the requirement that $x_{\rm rp}<x_{\rm Ni}$.

\subsection{Radioactive Heating} \label{subsec:heating}

The decay of $^{56}$Ni and $r$-process material heats up the ejecta and provides the energy source for SNe. The specific radioactive heating rate due to the decay chain $^{56}$Ni$\rightarrow^{56}$Co$\rightarrow^{56}$Fe can be written as \citep{Colgate1969,Colgate1980,Arnett1980,Arnett1982}
\begin{equation}
\dot{q}_{\mathrm{Ni}} =\left( \epsilon _{%
\mathrm{Ni}}-\epsilon _{\mathrm{Co}}\right) e^{-t/\tau _{\mathrm{Ni}%
}}+\epsilon _{\mathrm{Co}}e^{-t/\tau _{\mathrm{Co}}},
\end{equation}%
where $\epsilon _{\mathrm{Ni}%
}=3.9\times 10^{10}~\rm{erg}~\rm{g}^{-1}~\rm{s}^{-1}$, $\epsilon _{%
\mathrm{Co}}=6.78\times 10^{9}~\rm{erg}~\rm{g}^{-1}~\rm{s}^{-1}$. $\tau _{%
\mathrm{Ni}}=8.8~\rm days$ and $\tau _{\mathrm{Co}}=111.3~\rm days$ are the decay time of $^{56}$Ni decays to $^{56}$Co and $^{56}$Co decays to $^{56}$Fe, respectively. 

 The specific radioactive heating rate from the $r$-process material can be written as \citep{Metzger2010,Korobkin2012}
\begin{equation}
\dot{q}_{\rm rp}=4\times10^{18}\left[\frac{1}{2}-\frac{1}{\pi}{\rm
arctan}\left(\frac{t-t_0}{\sigma}\right)\right]^{1.3}~\rm
erg~s^{-1}~g^{-1}
\end{equation}
with $t_0=1.3$\,s and $\sigma=0.11$\,s.

We simply assume that the energy from the decay of $^{56}$Ni and $r$-process material heats the ejecta with 100\% efficiency. This assumption is roughly satisfied in the optically thick region. This may result in a slight underestimation of the mass of $^{56}$Ni in our model, but will not affect the mass of $r$-process material because the heating from $r$-process decay is secondary with respect to the heating due to the $^{56}$Ni decay chain in almost all stages of SN evolution \citep{Siegel2019}. For the optically thin region, this assumption would lead to an overestimation of the nebular luminosity. However, as discussed in Sections \ref{sec:results} and \ref{sec:discussion}, the uncertainty of the nebular emission has a minor influence on the results of this work.

\subsection{Opacity} \label{subsec:opacity}

A wavelength-independent gray opacity is assumed in our model. The opacity of any region in the ejecta depends on its composition and can be written as
\begin{equation}
    \kappa=\kappa_{\rm sn}(1-X_{\rm rp})+\kappa_{\rm rp}X_{\rm rp},
\end{equation}
where $\kappa_{\rm sn}$ is the opacity of the $r$-process-free ejecta, $X_{\rm rp}$ and $\kappa_{\rm rp}$ are the the mass fraction and the opacity of the $r$-process-rich outflows, respectively. $\kappa_{\rm rp}$ depends mainly on the electron fraction $Y_e$ of the outflows, which is quite uncertain considering the complexity of neutrino transport and angular momentum transport in numerical simulations. \cite{Siegel2019} found that for neutrino-cooled disks with sufficiently high accretion rates ($\sim0.003-0.1~M_\odot~\rm s^{-1}$), the electron fraction in the disk midplane (and then the outflows) can reach $Y_e\lesssim0.2$. Simulations employing different neutrino transport solutions show that the electron fractions of the outflows can be raised to $Y_e\gtrsim0.25$ \citep{Just2022a}, $Y_e\gtrsim0.3$ \citep{Miller2020}, and $Y_e\gtrsim0.4$ \citep{Fujibayashi2022}, respectively. The electron fraction determines the abundance of lanthanides and actinides that dominate the opacity of the mixture. \cite{Tanaka2020} found that the opacities for the mixture of $r$-process elements to be $\kappa_{\rm rp}\sim20-30~\rm cm^2~g^{-1}$ for $Y_{e} \lesssim 0.2$, $\kappa_{\rm rp}\sim3-5~\rm cm^2~g^{-1}$ for $Y_{e} \approx 0.25-0.35$, and $\kappa_{\rm rp}\sim1~\rm cm^2~g^{-1}$ for $Y_{e} \approx 0.4$ at temperature $T=5-10\times10^3~\rm K$. Therefore, as a result of the numerical simulations, $\kappa_{\rm rp}$ ranges roughly from $1$ to $30~\rm cm^2~g^{-1}$. \cite{Barnes2022} fixes the opacity of the ejecta without $^{56}$Ni and $r$-process material to $0.05~\rm cm^{2}~g^{-1}$, and expresses the opacity of pure $^{56}$Ni as a temperature-dependent segmentation function varying between $0.01~\rm cm^{2}~g^{-1}$ and $0.1~\rm cm^{2}~g^{-1}$. In this work, we set $\kappa_{\rm sn} = 0.05~\rm cm^{2}~g^{-1}$ by assuming that the two cases above have the same opacity. This difference is negligible considering the large opacity of the $r$-process material.

\subsection{Internal Energy Evolution and Emerging Radiation} \label{subsec:radiation}

We divide the homologously expanded ejecta into a series of concentric shell layers. The mass of each shell layer can be derived from the density profile in Section 2.1. The evolution of the internal energy of the $i$th layer is given by the following energy conservation equation
\begin{equation}
\frac{dE_{{\rm int},i}}{dt}=m_{{\rm Ni},i}\dot{q}_{{\rm Ni}}+m_{{\rm rp},i}\dot{q}_{{\rm rp}} -L_i-\frac{E_{{\rm int},i}}{r_i}\frac{dr_i}{dt},
\end{equation}%
where $E_{{\rm int},i}$ is the internal energy, $L_i$ is the luminosity of the net flow of energy from the $i$th layer, $m_{{\rm Ni},i}$ and $m_{{\rm rp},i}$ are the mass of $^{56}$Ni and $r$-process material, $r_i$ is the radius of the layer.

The output luminosity contributed by the $i$th layer can be estimated by
\begin{equation}
L_i=\frac{E_{{\rm int},i}}{\max \left[t_{\mathrm{d}, i}, t_{\mathrm{lc}, i}\right]}
\end{equation}
where $t_{\mathrm{d}, i}$ is the radiation diffusion timescale, $t_{\mathrm{lc}, i}=r_i/c$ limits the energy loss time to the light crossing time. The radiation diffusion timescale of the $i$th layer is given by
\begin{equation}
t_{\mathrm{d}, i}=\frac{3 \kappa}{4 \pi r_i c} \sum_{i^{\prime}=i}^n m_{i^{\prime}},
\end{equation}
where the effect of the shell outside the $i$th layer is considered by summing the masses of layers $i$ to $n$.
Finally, by summarising the contributions of all layers, the emerging bolometric luminosity of the entire ejecta can be obtained as follows
\begin{equation}
    L_{\rm bol}(t) = \sum\limits_i L_{i}.
\end{equation}

In optically thick regions, where matter is in local thermal equilibrium, radiation tends towards the blackbody spectrum. Assuming that all radiation in the optically thick regions is emitted from the photosphere in the form of a standard blackbody spectrum, one can define an effective temperature
\begin{equation}
T_{\rm eff}=\left(\frac{L_{\rm ph}}{4\pi\sigma_{\rm SB} R_{\rm ph}^2}\right)^{1/4},
\end{equation}
where $\sigma_{\rm SB}$ is the Stefan-Boltzmann constant, $L_{\rm ph}$ is the photospheric luminosity by summing the luminosity below the photospheric radius $R_{\rm ph}$. The photosphere is defined as the position where the optical depth is 2/3 from the outermost layer of the ejecta inwards, thus $R_{\rm ph}$ can be given by
\begin{equation}
\int_{R_{\text{ph}}}^{R_{\text{SN}}}\rho \kappa dr=\frac{2}{3},
\end{equation}
where $R_{\text{SN}}$ is the radius of the outermost layer of the ejecta. For homologously expanding ejecta, the photospheric velocity can be easily derived from
\begin{equation}
v_{\text{ph}}=\frac{v_{\text{SN}}R_{\text{ph}}}{R_{\text{SN}}}.
\end{equation}
The flux density of the photospheric emission at photon frequency $\nu$ can be given by
 \begin{equation}
F_{\nu}(t)=\frac{2\pi h\nu^3}{c^2}\frac{1}{\exp(h\nu/kT_{\rm
eff})-1}\frac{R_{\rm ph}^2}{D_{\rm L}^2},
\end{equation}
where $h$ is the Planck constant, $k$ is the Boltzmann constant, $D_{\rm L}=114$ Mpc is the luminosity distance of SN 2020bvc. 

\begin{figure}
\centering
\includegraphics[width=0.5\textwidth,angle=0]{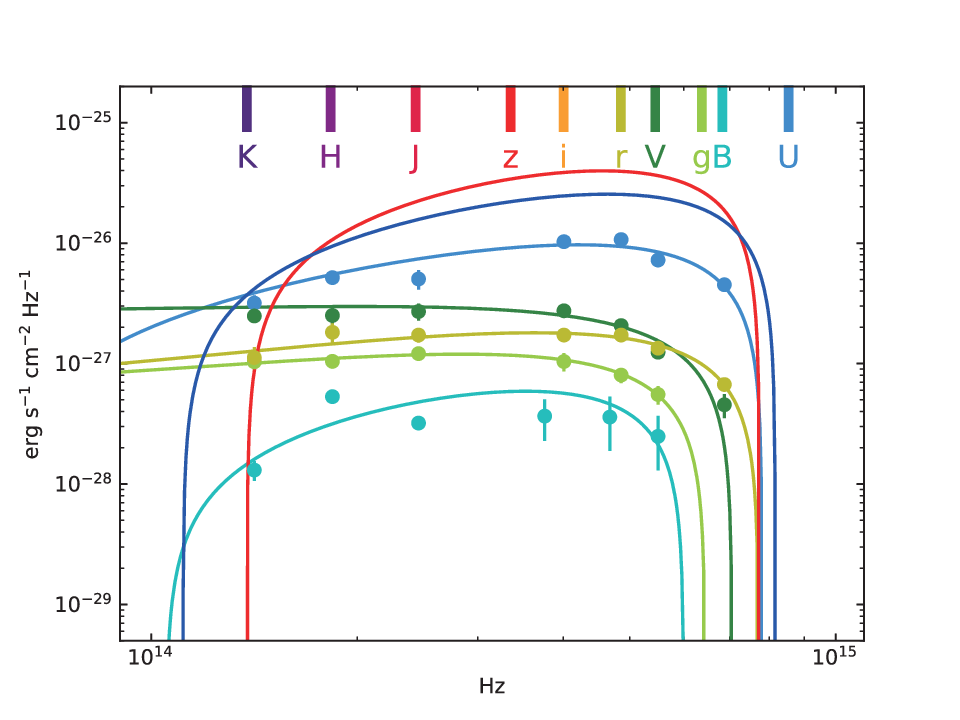}
\caption{The late-time photometry of SN 2007gr (Ic, 140 days and 380 days; \citealt{Hunter2009}), SN 2007C (Ib, 90 days; \citealt{Bianco2014}), SN 2007I (Ic-BL, 80 days; \citealt{Bianco2014}), SN 2007uy (Ib, 100 days; \citealt{Bianco2014}). We fit the photometry for each epoch with a parabola and the results are represented by solid lines in the corresponding colors. For comparison, we plot the scaled empirical $r$-process-free nebular SEDs derived from fitting the observations of SN 2020bvc based on $r$-process-enriched and $r$-process-free models, represented by the red and blue solid lines, respectively.
\label{fig:nebSED}}
\end{figure}

The nebular emission of ejecta mixed with $r$-process material from optically thin regions is highly uncertain. We follow \cite{Barnes2022} in treating nebular emission as a combination of of two different spectral energy distributions (SEDs) associated with $r$-process and $r$-process-free material. The blackbody spectrum at a temperature of 2500 K is treated as an $r$-process SED. However, for the $r$-process-free material, we loose the limitation on the $r$-process-free SED and express it as an empirical quadratic function
\begin{equation}
\mathcal{F}(\nu)=-f_a(\nu-f_h)^2+f_k,
\end{equation}
where $f_a,f_h,f_k$ are positive coefficients. As shown in Figure \ref{fig:nebSED}, the empirical function can roughly reproduce the late photometry of type Ib/c SNe.

Finally, we determine the monochromatic AB magnitude by $M_{\nu}=-2.5\log_{10}(F_{\nu}/3631\rm
Jy)$.

\subsection{Shock Cooling Emission} \label{subsec:cooling}

In order to explain the first peak of SN 2020bvc, we introduced the shock cooling emission in our model. We assume that the progenitor was surrounded by extended material with mass $M_e$ and radius $R_e$. As the SN shock powered by the explosion energy $E_{\rm SN}$ moves from the ejecta to the extended material, it will transition to a new velocity $v_e$ which can be expressed in the following \citep{Nakar2014}
\begin{equation}
v_e \approx 1.5 \times 10^9E_{\rm SN,51}^{0.5}\left(\frac{M_{\mathrm{ej}}}{3 M_{\odot}}\right)^{-0.35}\left(\frac{M_{\mathrm{e}}}{0.01 M_{\odot}}\right)^{-0.15} \mathrm{~cm} \mathrm{~s}^{-1},
\end{equation}
where $E_{\rm SN,51}=E_{\rm SN}/10^{51}$ erg, then the energy carried by the extended material is
\begin{equation}
E_e \approx 2 \times 10^{49} E_{\rm SN,51}\left(\frac{M_{\mathrm{ej}}}{3 M_{\odot}}\right)^{-0.7}\left(\frac{M_{\mathrm{e}}}{0.01 M_{\odot}}\right)^{0.7} \mathrm{erg}.
\end{equation}
Assuming that the extended material has a uniform density profile $\rho_e(t)=3 M_{\mathrm{e}} / 4 \pi R(t)^3$, where $R(t)=R_{\mathrm{e}}+v_{\mathrm{e}} t$,  the luminosiy of shock cooling emission can be expressed analytically as \citep{Piro2015}
\begin{equation}
L_e(t)=\frac{t_e E_e}{t_p^2} \exp \left[-\frac{t\left(t+2 t_e\right)}{2 t_p^2}\right],
\end{equation}
where $t_p$ is the diffusion timescale and can be roughly estimated as \citep{Arnett1982}
\begin{equation}
t_p \approx\left(\frac{M_e \kappa_e}{v_e c}\right)^{1 / 2},
\end{equation}
where $\kappa_e$ is the opacity of the extented material, which we set equal to the opacity of the ejecta without $r$-process material, i.e., $\kappa_e=\kappa_{\rm sn}=0.05~\rm cm^{2}~g^{-1}$.

\section{Fitting and results} \label{sec:results}

\begin{figure}
\centering
\includegraphics[width=0.49\textwidth,angle=0]{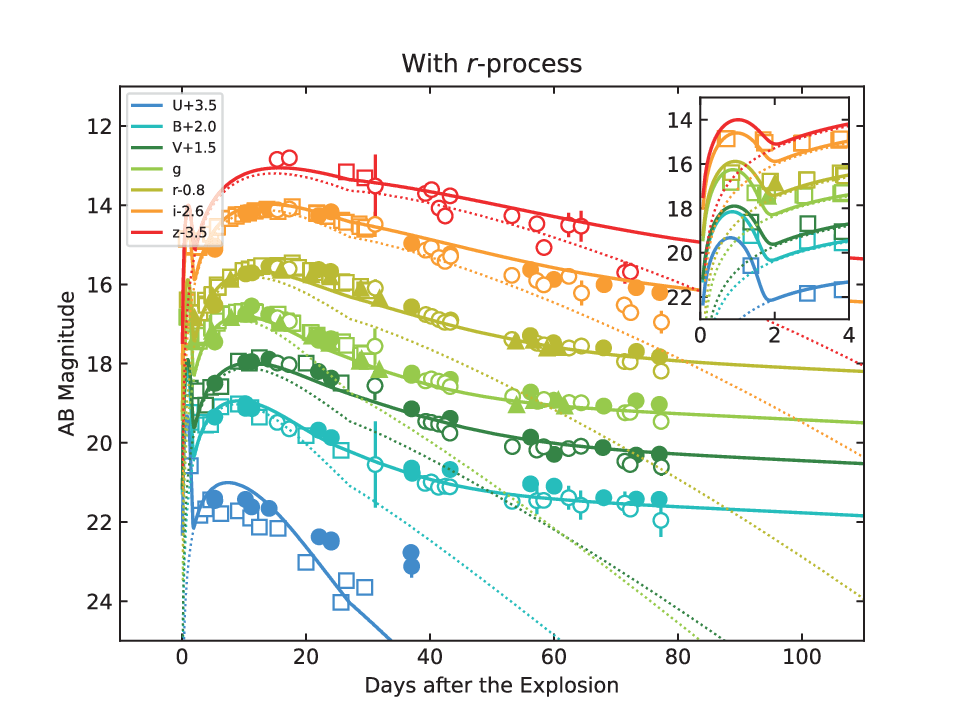}
\includegraphics[width=0.49\textwidth,angle=0]{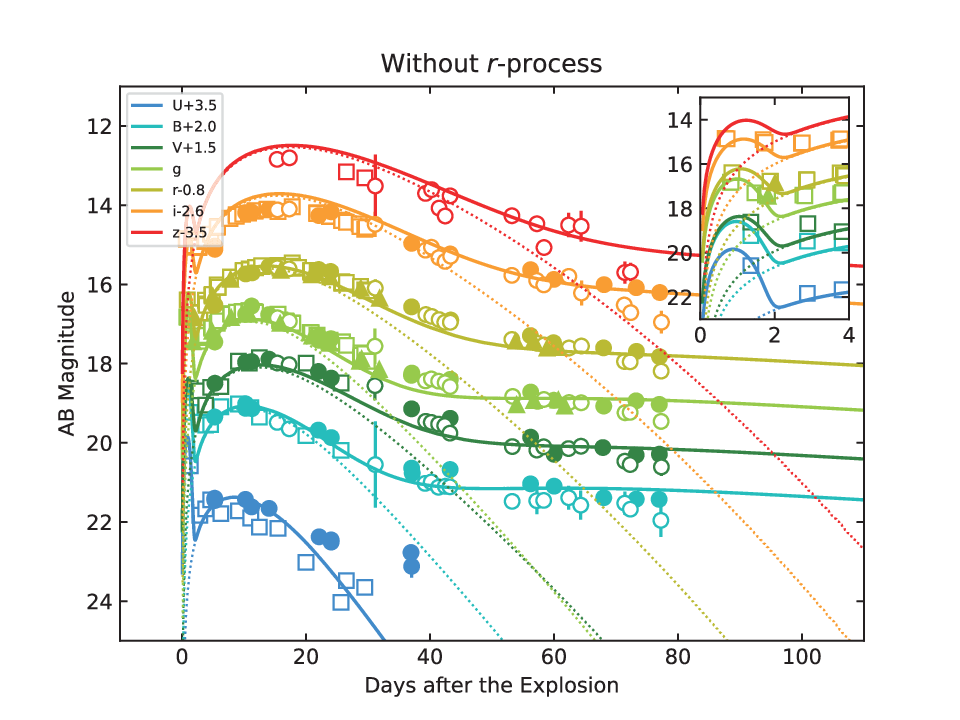}
\caption{Multi-band light curves of SN 2020bvc and the best-fitting results for SN models with (left) and without (right) $r$-process material. Data for LCO (filled circles), Konkoly (open circles), and ZTF (filled Triangles) are from Rho et al. 2021. Early photometric data from Ho et al. 2020 are represented by open squares. 
The time zero point is set to the most recent nondetection before the first detection (MJD = 58,882.67; Ho et al. 2020).
The solid and dotted lines represent the total and photospheric emission, respectively.
\label{fig:multi}}
\end{figure}

\begin{figure}
\centering
\includegraphics[width=0.49\textwidth,angle=0]{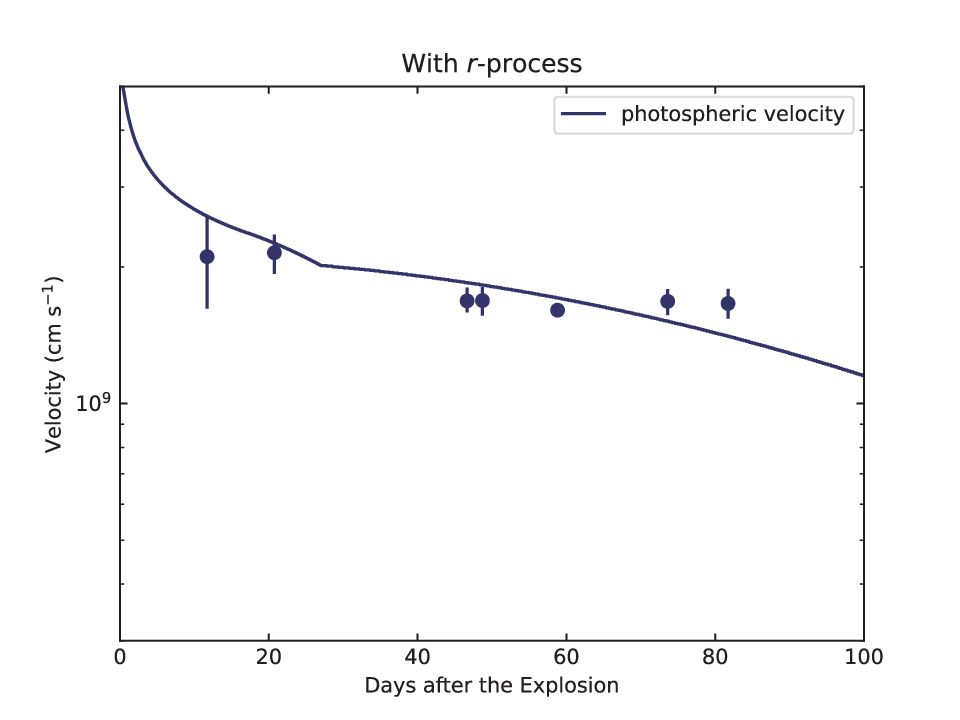}
\includegraphics[width=0.49\textwidth,angle=0]{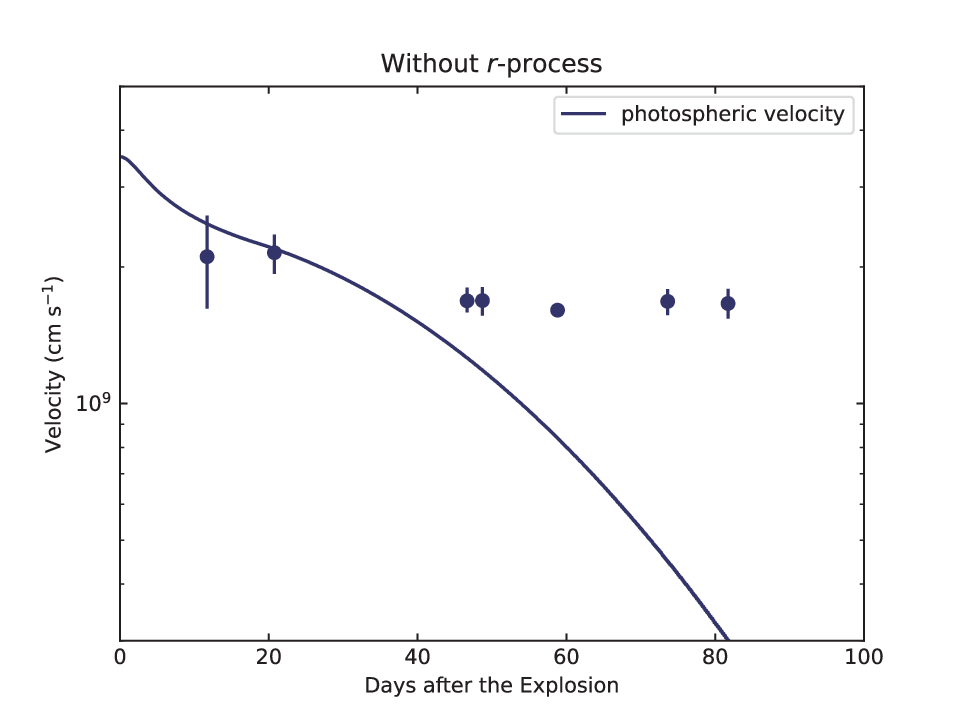}
\caption{Fe II $\lambda$5169 absorption velocity evolution of SN 2020bvc and the best-fitting photospheric velocity evolutions for SN models with (left) and without (right) $r$-process material. Data are taken from Rho et al. 2021.
\label{fig:vel}}
\end{figure}

The multi-band light curves of SN 2020bvc are taken from \cite{Ho2020b} and \cite{Rho2021}, where the early double-peaked light curves are mainly contributed by \cite{Ho2020b}. The $UBVgriz$-band data used for the fitting are all presented in the AB magnitude system (the $UBV$-band data from \cite{Rho2021} have been calibrated from Vega magnitudes to AB magnitudes), as shown in Figure \ref{fig:multi}.  In addition, the Fe II $\lambda$5169 absorption velocities of SN 2020bvc from \cite{Rho2021} are regarded as part of the data used for the fitting\footnote{Here we consider the Fe II $\lambda$5169 line velocities of SN 2020bvc as photospheric velocities (e.g., \citealt{Branch2002}; for a caveat see \citealt{Dessart2015}.}, as shown in Figure \ref{fig:vel}.

We develop a Fortran-based numerical model based on the description in Section \ref{sec:model} and implement the Markov Chain Monte Carlo (MCMC) techniques by using the \texttt{emcee} Python package \citep{emcee}. \texttt{F2PY} is employed to provide a connection between Python and Fortran languages. We perform the MCMC with 26 walkers for running at least 100,000 steps, until the step is longer than 50 times the integrated autocorrelation time $\tau_{\rm ac}$, i.e. $N_{\rm step}=\max(10^5, 50\tau_{\rm ac})$ to make sure that the fitting is sufficiently converged \citep{emcee}. Once the MCMC is done, the best-fitting values and the $1\sigma$ uncertainties are computed as the 50th, 16th, and 84th percentiles of the posterior samples.

The free parameters required for the fitting include: the total kinetic energy of SN ejecta ($E_{\rm SN}$), the total mass of SN ejecta ($M_{\rm ej}$), the mass of radioactive $^{56}$Ni ($M_{\rm Ni}$), the mass of $r$-process material ($M_{\rm rp}$), the opacity of $r$-process material ($\kappa_{\rm rp}$), the dimensionless radius of the break in the SN ejecta density profile from the inner component to the outer component ($x_0$), the maximum dimensionless mixing radius of $^{56}$Ni ($x_{\rm Ni}$), the maximum dimensionless mixing radius of $r$-process material ($x_{\rm rp}$), the SN explosion time ($t_{\rm shift}$), the total mass of extended material ($M_e$), the maximum radius of extended material ($R_e$), the coefficients of the empirical function of the $r$-process-free nebular SED ($f_a$, $f_h$, and $f_k$). Uniform priors on $E_{\rm SN}/10^{51}~\rm erg$, $M_{\rm ej}/M_\odot$, $M_{\rm Ni}/M_\odot$, $M_{\rm rp}/M_\odot$, $\kappa_{\rm rp}/{\rm cm^2~g^{-1}}$, $x_0$, $x_{\rm Ni}$, $x_{\rm rp}$, $t_{\rm shift}/\rm day$, $M_e/M_\odot$, $R_e/R_\odot$, $\log f_a$, $\log f_h$, and $\log f_k$ are adopted in this paper. In order to explore a parameter space as large as possible, we set a wide enough range for the priors. As a comparison, we also perform a fitting based on a $r$-process-free model that simply makes $M_{\rm rp}=0$. The free parameters and priors in our model can be seen in Table \ref{tab:pars}.

Figure \ref{fig:multi} shows the best-fitting multi-band light curves based on the $r$-process-enriched model and the $r$-process-free model. We find that for both models the overall quality of the fitting is good, except for a few photometric data that deviate slightly from the best-fitting light curves. This deviation is mainly caused by the inconsistency of the photometric data. After showing the photometric data from different telescopes with different markers, one can find a significant offset between the late-time $BVgri$-band light curves from LCO (filled circles) and that from Konkoly (open circles). The late-time $z$-band light curve from Konkoly also shows two different trends. Due to the difference in filters, there is also an inconsistency of $U$-band photometry between LCO from \cite{Rho2021} and Swift/UVOT from \cite{Ho2020b}. Figure \ref{fig:vel} shows the best-fitting photospheric velocity evolution based on both models. The best-fitting photospheric velocities from the $r$-process-enriched model matches well with the Fe II $\lambda$5169 absorption velocities of SN 2020bvc, while those from $r$-process-free model drop off so rapidly, in contrast to the almost identical high Fe II velocities at late times.

Table \ref{tab:pars} shows the best-fitting parameters and their corresponding 1$\sigma$ uncertainties from both models. The reduced $\chi^2$ derived by both models are large. $\chi^2_{\rm tot}/{\rm dof}\approx55$ for $r$-process-enriched model is slightly smaller than $\chi^2_{\rm tot}/{\rm dof}\approx59$ for $r$-process-free model. 
When only counting multi-band light curves, $\chi^2_{\rm lc}/{\rm dof}\approx56$ and $\chi^2_{\rm lc}/{\rm dof}\approx59$ for $r$-process-enriched and $r$-process-free models. When only considering velocity evolution, $\chi^2_{\rm vel}/{\rm dof}\approx3$ and $\chi^2_{\rm vel}/{\rm dof}\approx70$ for $r$-process-enriched and $r$-process-free models, respectively. This suggests that the $r$-process-enriched model has almost no advantage over the $r$-process-free model in explaining the multi-band light curves of SN 2020bvc, but the $r$-process-enriched model is significantly better than the $r$-process-free model in explaining the velocity evolution. To verify whether the large $\chi^2$ is related to the dispersion of the data, we divided the dataset into two groups based on the evolution of the multi-band light curves: the flatter evolving dataset A1, and the steeper evolving dataset A2, as shown in Figure \ref{fig:multi2} in Appendix \ref{sec:datasets}. After fitting the $r$-process-enriched model to data sets A1 and A2, we derived $\chi^2_{\rm tot}/{\rm dof}\approx45$ and $\chi^2_{\rm tot}/{\rm dof}\approx58$, respectively. This suggests that the $r$-process-enriched model is more compatible with data set A1, and that the dispersion of the data is only responsible for a fraction of the large $\chi^2$.

The best-fitting results from both models are similar. We find that in order to explain the main peak requires a total kinetic energy of $E_{\rm SN}\approx2\times10^{52}~\rm erg$, a total ejecta mass of $M_{\rm ej}\approx5.7-5.9~M_\odot$, and a radioactive nickel mass of $M_{\rm Ni}\approx0.6-0.8~M_\odot$. The maximum dimensionless mixing radius $x_{\rm Ni}\approx 1.0$ suggests that the $^{56}$Ni is almost completely mixed with the entire ejecta. In order to explain the first peak of the light curves, an extended material with $M_e\approx0.05~M_\odot$, $R_e\approx130-180~R_\odot$ is needed. The estimated total mass of the extended material is lower than other type Ic SNe (e.g., \citealt{Piro2015,Taddia2018}), but optically thick enough to ensure continuous propagation of the shock (e.g. Equation 1 from \citealt{Piro2015}). In Figure \ref{fig:nebSED}, we plot the best-fitting $r$-process-free nebular SEDs as well as the late-time photometry of several SNe. The spectral shape are similar to those of the late-time photometry of type Ib/c SNe, with a quite flat profile between the $B$ and $z$ bands. For $r$-process-enriched model, the intermediate opacity $\kappa_{\rm rp}\approx4~\rm cm^2~g^{-1}$ derived from the fitting results suggests that the outflows may have only undergone the $r$-process extending to the second abundance peak ($A\sim130$), with a small or zero mass fraction of lanthanides. A total $r$-process mass of $M_{\rm rp}\approx0.36~M_\odot$ and a maximum mixing radius of $x_{\rm rp}\approx0.5$ are needed to account for the photospheric velocity evolution. The velocity evolution of the $r$-process-enriched model is comparable to that of the $r$-process-free model until about 30 days, when the $r$-process material is still deeply buried within the photosphere. Thereafter, $r$-process material begins to emerge, slowing the recession of the photosphere due to its great opacity, keeping the photoshperic velocity at a high level for a long time. In Appendix \ref{sec:corner}, we display the corner plots showing the results of our MCMC parameter estimation. We note that several parameters show degeneracies, as it can be seen from the correlations in Figure \ref{fig:cornerA} and \ref{fig:cornerB}. This means that there are many combinations of parameters that are compatible with the observations. In particular, there is a strong anti-correlation between $M_{\rm rp}$ and $\kappa_{\rm rp}$. It can be expected that outflows with high opacity/low mass or low opacity/high mass combinations can also explain the observed multi-band light curves and velocity evolution.

\begin{figure}
\centering
\includegraphics[width=0.49\textwidth,angle=0]{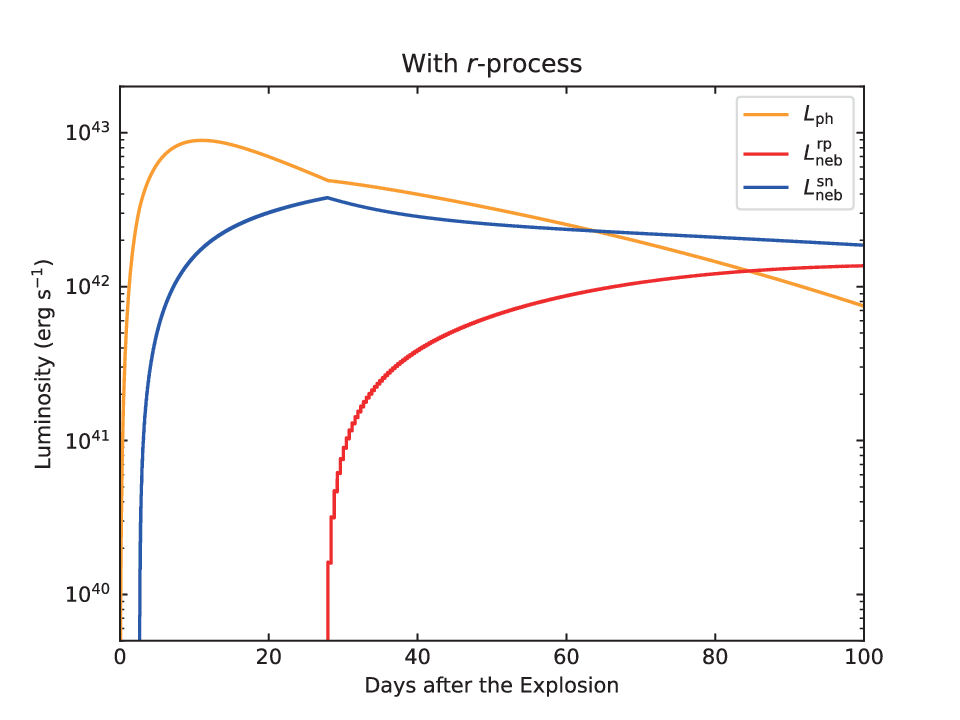}
\includegraphics[width=0.49\textwidth,angle=0]{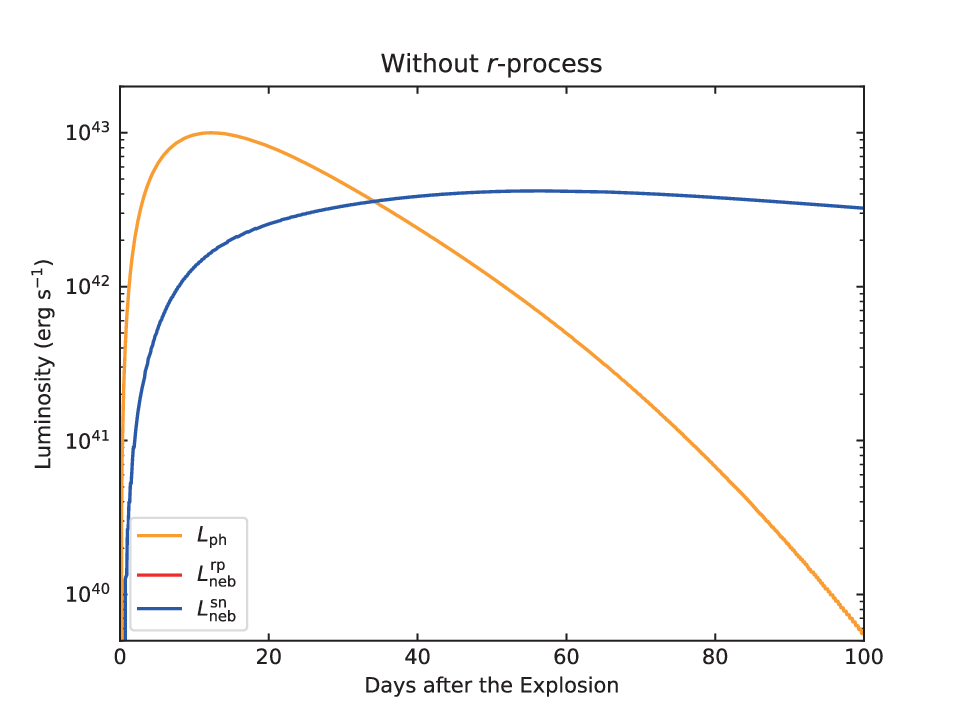}
\caption{Photospheric luminosity ($L_{\rm ph}$), $r$-process nebular luminosity ($L_{\rm neb}^{\rm rp}$), and $r$-process-free nebular luminosity ($L_{\rm neb}^{\rm sn}$) evolutions dervied from the best-fitting results for SN models with (left) and without (right) $r$-process material.
\label{fig:lum}}
\end{figure}

In Figure \ref{fig:lum}, we plot the luminosity evolution of the different components. 
The mixing of the $r$-process material redistributes the energy between the luminosity components. The late-time photosphere luminosity $L_{\rm ph}$ is comparable to the $r$-process-free nebular luminosity $L_{\rm neb}^{\rm sn}$, in contrast to the late-time dominance of $L_{\rm neb}^{\rm sn}$ in the $r$-process-free model. Within 80 days, however, the contribution of the $r$-process nebular luminosity $L_{\rm neb}^{\rm rp}$ is not dominant. This implies that the uncertainty in the $r$-process nebular SED has little effect on the fitting results.

\begin{deluxetable}{lcrcr}\label{tab:pars}
\tabletypesize{\footnotesize}
\tablecaption{Free parameters, priors, and best-fitting results in our model.}
\tablehead{
\colhead{Parameter}	&	\colhead{Prior A}	&	\colhead{Result A}	&	\colhead{Prior B}	&	\colhead{Result B}	\\
\colhead{} & \multicolumn{2}{c}{(with $r$-process)}			&	\multicolumn{2}{c}{(without $r$-process)}			
}
\startdata
$E_{\rm SN}~(10^{51}~{\rm erg})$	&	$[0,100]$	&	$21.10^{+1.44}_{-1.81}$	&	$[0,100]$	&	$20.43^{+2.02}_{-2.25}$	\\
$M_{\rm ej}~(M_\odot)$	&	$[0,10]$	&	$5.88^{+0.21}_{-0.27}$	&	$[0,10]$	&	$5.72^{+0.66}_{-0.75}$	\\
$M_{\rm Ni}~(M_\odot)$	&	$[0,1]$	&	$0.77^{+0.03}_{-0.04}$	&	$[0,1]$	&	$0.61^{+0.03}_{-0.03}$	\\
$M_{\rm rp}~(M_\odot)$	&	$[0,1]$	&	$0.36^{+0.26}_{-0.21}$	&	-	&	-	\\
$\kappa_{\rm rp}~(\rm cm^{2}~g^{-1})$	&	$[1,30]$	&	$4.29^{+2.09}_{-2.28}$	&	-	&	-	\\
$x_0$	&	$[0,1]$	&	$0.66^{+0.00}_{-0.00}$	&	$[0,1]$	&	$0.64^{+0.00}_{-0.00}$	\\
$x_{\rm Ni}$	&	$[0,1]$	&	$0.98^{+0.01}_{-0.01}$	&	$[0,1]$	&	$0.97^{+0.01}_{-0.01}$	\\
$x_{\rm rp}$	&	$[0,1]$	&	$0.54^{+0.03}_{-0.02}$	&	-	&	-	\\
$t_{\rm shift}~(\rm day)$	&	$[0,0.67]$	&	$0.00^{+0.00}_{-0.00}$	&	$[0,0.67]$	&	$0.00^{+0.00}_{-0.00}$	\\
$M_e~(M_\odot)$	&	$[0,1]$	&	$0.05^{+0.02}_{-0.01}$	&	$[0,1]$	&	$0.05^{+0.02}_{-0.03}$	\\
$R_e~(R_\odot)$	&	$[50,5000]$	&	$180.51^{+49.56}_{-59.77}$	&	$[50,5000]$	&	$130.96^{+65.22}_{-29.05}$	\\
$\log[f_a]$	&	$[-50,0]$	&	$-44.12^{+0.07}_{-0.06}$	&	$[-50,0]$	&	$-44.39^{+0.04}_{-0.03}$	\\
$\log[f_h]$	&	$[13,15]$	&	$14.66^{+0.01}_{-0.01}$	&	$[13,15]$	&	$14.67^{+0.01}_{-0.01}$	\\
$\log[f_k]$	&	$[-20,0]$	&	$-15.01^{+0.04}_{-0.04}$	&	$[-20,0]$	&	$-15.29^{+0.03}_{-0.02}$	\\
\hline									
$\chi^2_{\rm lc}/\rm{dof}$	&	-	&	$20533/365$	&	-	&	$21775/368$	\\
$\chi^2_{\rm vel}/\rm{dof}$	&	-	&	$21/7$	&	-	&	$492/7$	\\
$\chi^2_{\rm tot}/\rm{dof}$	&	-	&	$20554/372$	&	-	&	$22267/375$	\\
\enddata
\tablecomments{The uncertainties of the best-fitting parameters are measured as $1\sigma$ confidence ranges.}
\end{deluxetable}

\section{Discussion \& Conclusions} \label{sec:discussion}

\begin{figure}
\centering
\includegraphics[width=0.5\textwidth,angle=0]{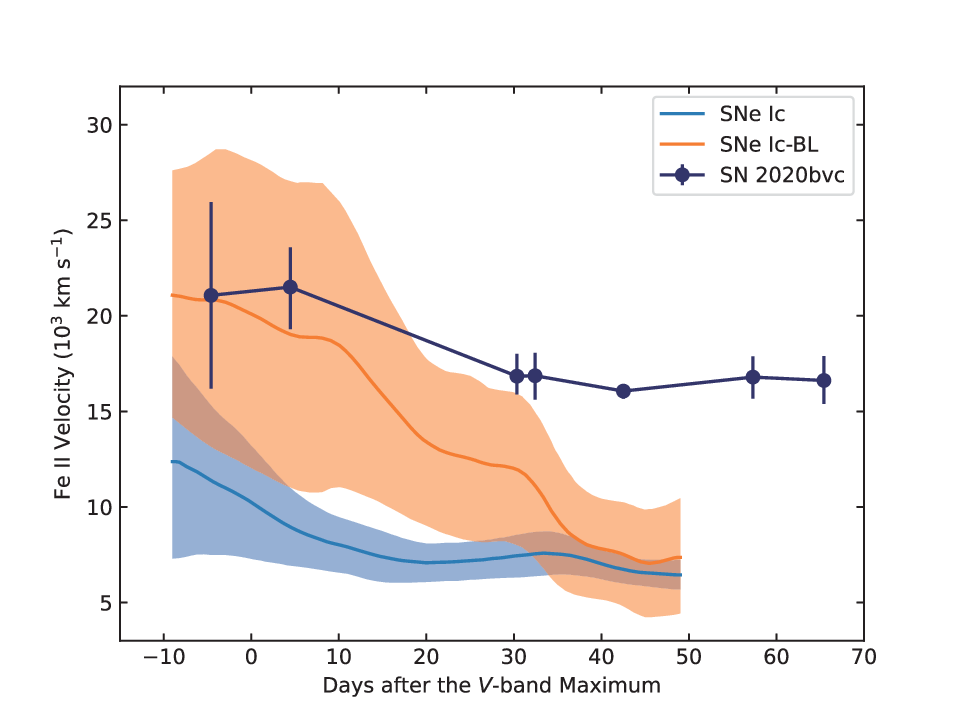}
\caption{Fe II $\lambda5169$ absorption velocities of SN 2020bvc (black) compared to the weighted moving average velocities of SNe Ic (blue) and SNe Ic-BL (orange) from \cite{Liu2017}.
\label{fig:compare}}
\end{figure}

In Figure \ref{fig:compare}, we compare the Fe II $\lambda$5169 absorption velocity evolution of SN 2020bvc with a large sample of SNe Ic and SNe Ic-BL \citep{Modjaz2016,Liu2017}. The early high velocities of SN 2020bvc are well consistent with those of SNe Ic-BL, but significantly higher than those of SNe Ic. At the $V$-band maximum, the weighted average absorption velocities in SNe Ic-BL and SNe Ic are $18,500\pm7400~\rm km~s^{-1}$ and $8000\pm1400~\rm km~s^{-1}$ \citep{Liu2017}, respectively, while the absorption velocity for SN 2020bvc is $\sim21,000~\rm km~s^{-1}$. Thereafter, the average velocities of SNe Ic/Ic-BL continued to decrease to $\lesssim10,000~\rm km~s^{-1}$, which contrasts with the unusually slow decreasing velocities of 2020bvc. The slow decay of the velocities of SN 2020bvc is followed by a long plateau phase, which has also been observed in other SNe Ic/Ic-BL, but the plateau of SN 2020bvc is much higher ($\sim16,000~\rm km~s^{-1}$). Previous radiation transfer simulations of stripped-envelope SNe indeed show a late flat evolution of the photospheric velocity, which is similar to the observed velocity evolution of SNe Ic/Ic-BL and may be attributed to recombination of ejecta and/or other radiative transfer effects (e.g. \citealt{Dessart2012,Moriya2020}). However, modelling such a high velocity plateau for SN 2020bvc based on radiative transfer models is a challenge, and a wide parameter space needs to be explored to determine whether observations can be interpreted without invoking the $r$-process.

The largest uncertainty in our emission model arises from the uncertainty in the $r$-process-free nebular SED. \cite{Barnes2022} fitted the photometric data from $t\approx120$ days after B-band maximum of type Ic SN 2007gr with a spline and adopted it as the $r$-process-free nebular SED. However, it is questionable whether the late-time SED of SN 2007gr can act as a representative of standard $r$-process-free SNe. As shown in Figure \ref{fig:nebSED}, the late-time photometry of type Ib/c SNe exhibits diversity. Even for the same SN (SN 2007gr), the late-time photometry is different at different epochs. We therefore loosen the restrictions on the $r$-process-free nebular SED by including an empirical function with three free parameters in our fitting. Nevertheless the $r$-process-free nebular SED will only have an effect on the late-time multi-band light curves, not on the early-time light curves and the photospheric velocity evolution. This is crucial as the properties of the explosion and ejecta are mainly determined by the main peak and the early-time photospheric velocity. This conclusion is supported by the fact that the best-fitting parameters derived from both models are almost identical, although their $r$-process-free nebular SEDs are not the same. Therefore, the uncertainty in the nebular SED has little impact on the properties of ejecta and explosions. On the other hand, the photospheric velocity is mainly determined by the photospheric location, which in turn is governed by the density and opacity of the ejecta. The density profile of the ejecta is related to the total mass of the ejecta. As the fitting result for $r$-process-free model shows, for a certain ejecta mass (determined by the early-time light curves), ejecta with typical density profile and opacity cannot reproduce the late-time photospheric velocity evolution of SN 2020bvc. The plausible explanation for the consistently high photosperic velocity is that the exceptionally high opacity of the material in the inner layers of the ejecta prevents the recession of the photosphere. In our model, this high opacity material originates from the neutron-rich disk outflows of the collaspar that have undergone a $r$-process nucleosynthesis. The derived opacity of $r$-process material $\kappa_{\rm rp}\approx4~\rm cm^2~g^{-1}$ suggests the electron fraction of the outflows to be $Y_{e} \approx 0.25-0.35$. The required mass of $r$-process material $M_{\rm rp}\approx0.36~M_\odot$ suggests a total accreted mass of $\gtrsim1.2~M_\odot$ based on the numerical simulations \citep{Siegel2017,Siegel2018,Fernandez2019}. In addition, the possible association of SN 2020bvc with a low luminosity GRB also hints at the origin of the collapsar.

In this paper, we investigate the multi-band light curves and photospheric velocity evolution of SN 2020bvc based on a semi-analytic emission model with and without $r$-process matter. Although the $r$-process-enriched model has no advantage over the $r$-process-free model in explaining the multi-band light curves of SN 2020bvc, the $r$-process-enriched model can reproduce the photospheric velocity evolution significantly better than the $r$-process-free model. The fitting results are robust and weakly dependent on the $r$-process and $r$-process-free nebular SEDs. We therefore conclude that SN2020bvc is likely to have originates from a collapsar. A moderate amount of $r$-process material was produced in the disk outflows of this collapsar and partially mixed with the ejecta to significantly reshape the photospheric velocity evolution of SN 2020bvc. Future statistical analysis of a sample of type Ic-BL SNe with both multi-band photometry and rich and detailed spectroscopic information will help us to understand how much collapsars can contribute to the abundance of the $r$-process abundance in the Universe.

\begin{acknowledgments}
We thank the anonymous referee for the helpful suggestions that allowed us to improve our manuscript. L.L. and S.Q.Z. acknowledge support from the National Natural Science Foundation of China (grant No. 12247144) and China Postdoctoral Science Foundation (grant Nos. 2021TQ0325 and 2022M723060). Z.G.D. is supported by the National Key Research and Development Program of China (grant No. 2017YFA0402600), National SKA Program of China (grant No. 2020SKA0120300), and National Natural Science Foundation of China (grant No. 11833003).
\end{acknowledgments}

\vspace{5mm}
\software{emcee \citep{emcee}}
\clearpage

\appendix

\section{Best-fitting results for data sets A1 and A2} \label{sec:datasets}

\begin{figure}[b]
\centering
\includegraphics[width=0.49\textwidth,angle=0]{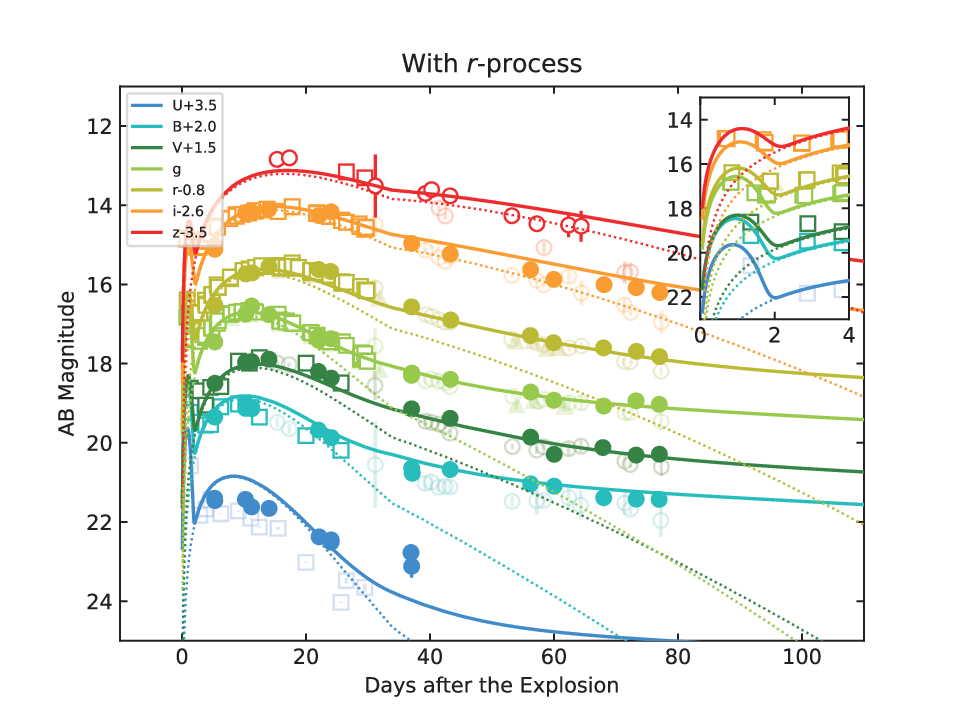}
\includegraphics[width=0.49\textwidth,angle=0]{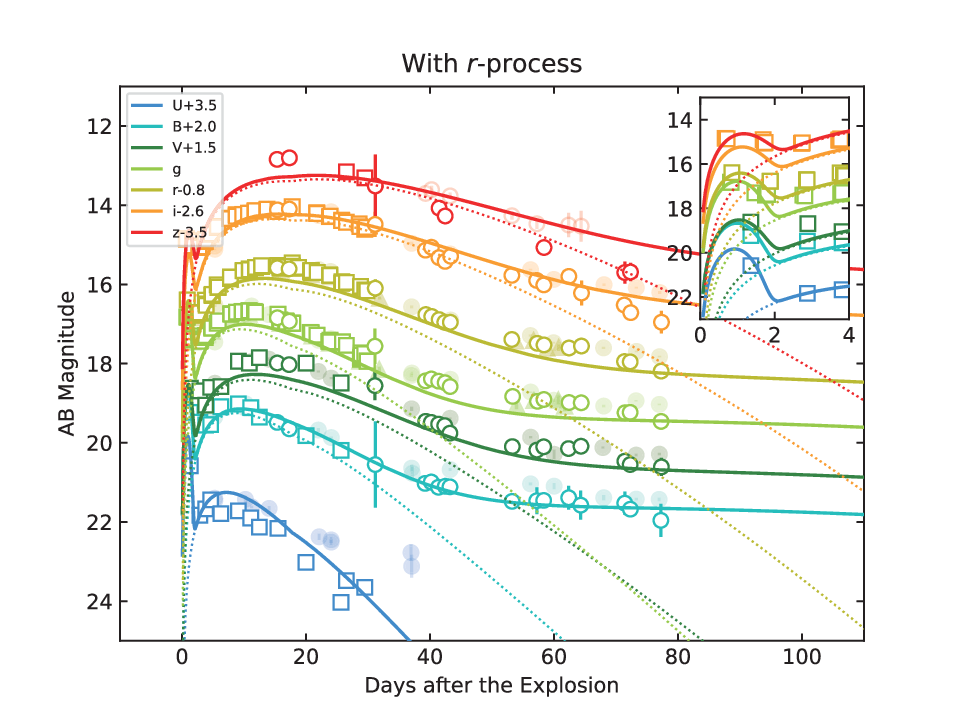}
\caption{Multi-band light curves for 2020bvc from data sets A1 (left) and A2 (right), and the best-fitting results based on $r$-process-enriched model. Data for LCO (filled circles), Konkoly (open circles), and ZTF (filled Triangles) are from Rho et al. 2021. Early photometric data from Ho et al. 2020 are represented by open squares. 
The time zero point is set to the most recent nondetection before the first detection (MJD = 58,882.67; Ho et al. 2020).
The solid and dotted lines represent the total and photospheric emission, respectively.
\label{fig:multi2}}
\end{figure}

\begin{figure}[b]
\centering
\includegraphics[width=0.49\textwidth,angle=0]{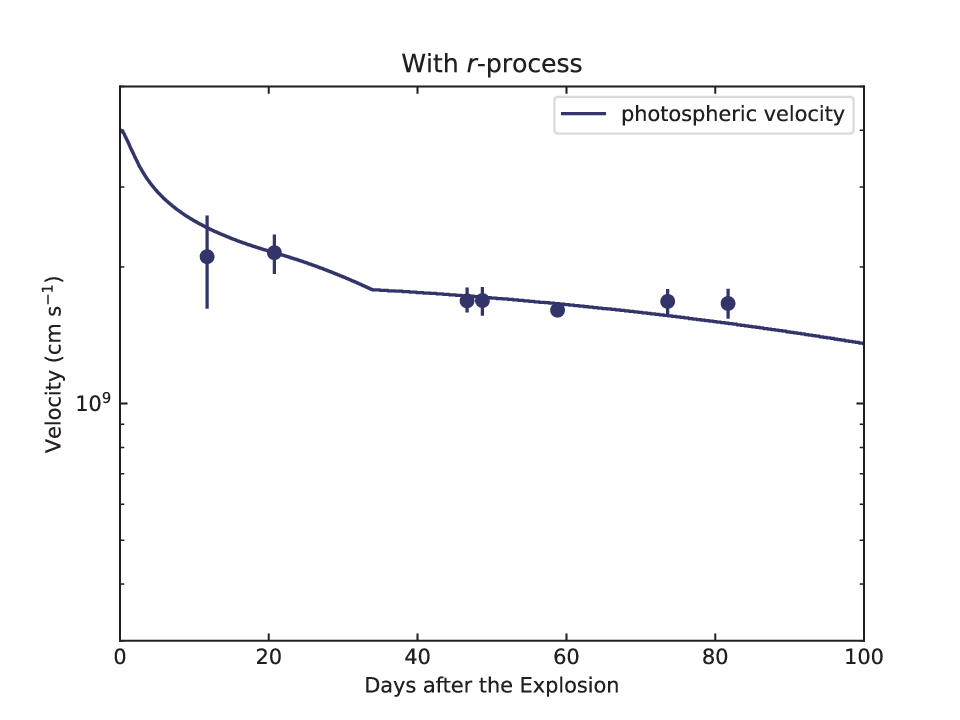}
\includegraphics[width=0.49\textwidth,angle=0]{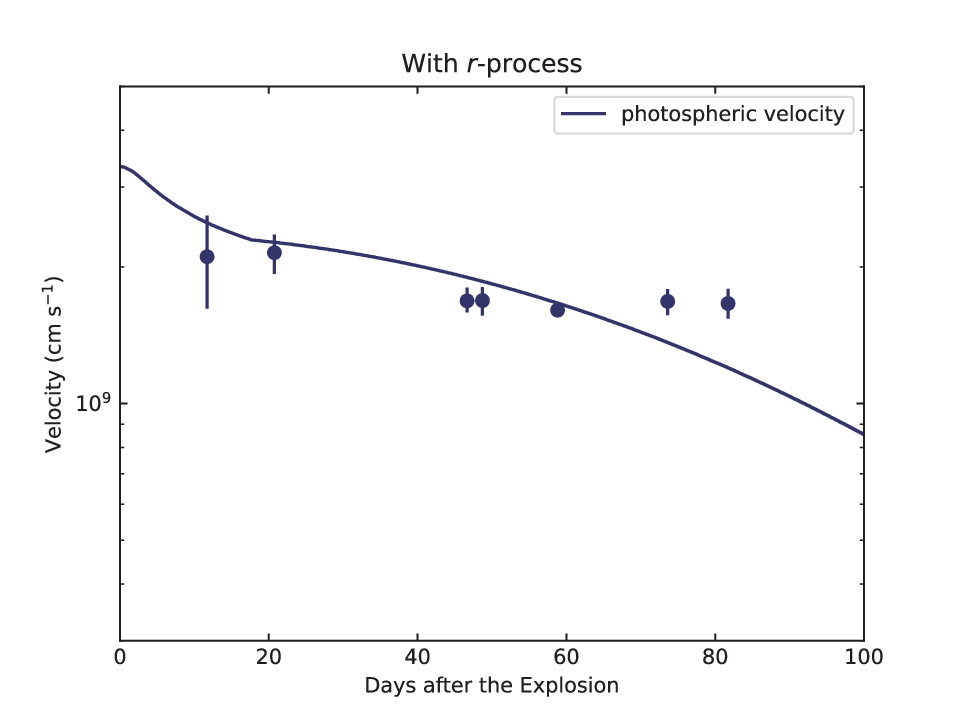}
\caption{Fe II $\lambda$5169 absorption velocity evolution of SN 2020bvc and the best-fitting photospheric velocity evolution derived from fitting the $r$-process-enriched model to data sets A1 (left) and A2 (right). Data are taken from Rho et al. 2021.
\label{fig:vel2}}
\end{figure}

We show the best-fitting multi-band light curves and photospheric velocity evolution derived from fitting the $r$-process-enriched model to data sets A1/A2 in Figures \ref{fig:multi2} and \ref{fig:vel2}, respectively. Table \ref{tab:pars2} shows the corresponding best-fitting parameters and their $1\sigma$ uncertainties.

\begin{deluxetable}{lcrr}\label{tab:pars2}
\tablecaption{Free parameters, priors, and best-fitting results in our model.}
\tablehead{
\colhead{Parameter}	&	\colhead{Prior A}	&	\colhead{Result A1}	&	\colhead{Result A2}	\\
}
\startdata
$E_{\rm SN}~(10^{51}~{\rm erg})$	&	$[0,100]$	&	$20.58^{+1.49}_{-1.78}$	&	$21.93^{+2.20}_{-2.79}$	\\
$M_{\rm ej}~(M_\odot)$	&	$[0,10]$	&	$5.33^{+0.26}_{-0.26}$	&	$5.89^{+0.53}_{-0.15}$	\\
$M_{\rm Ni}~(M_\odot)$	&	$[0,1]$	&	$0.70^{+0.02}_{-0.03}$	&	$0.74^{+0.02}_{-0.02}$	\\
$M_{\rm rp}~(M_\odot)$	&	$[0,1]$	&	$0.21^{+0.19}_{-0.11}$	&	$0.15^{+0.07}_{-0.07}$	\\
$\kappa_{\rm rp}~(\rm cm^{2}~g^{-1})$	&	$[1,30]$	&	$4.26^{+2.26}_{-2.35}$	&	$5.98^{+1.05}_{-1.31}$	\\
$x_0$	&	$[0,1]$	&	$0.61^{+0.01}_{-0.00}$	&	$0.44^{+0.00}_{-0.00}$	\\
$x_{\rm Ni}$	&	$[0,1]$	&	$0.99^{+0.01}_{-0.01}$	&	$0.98^{+0.01}_{-0.02}$	\\
$x_{\rm rp}$	&	$[0,1]$	&	$0.50^{+0.03}_{-0.02}$	&	$0.40^{+0.01}_{-0.02}$	\\
$t_{\rm shift}~(\rm day)$	&	$[0,0.67]$	&	$0.00^{+0.00}_{-0.00}$	&	$0.00^{+0.00}_{-0.00}$	\\
$M_e~(M_\odot)$	&	$[0,1]$	&	$0.05^{+0.01}_{-0.00}$	&	$0.04^{+0.01}_{-0.01}$	\\
$R_e~(R_\odot)$	&	$[50,5000]$	&	$130.32^{+60.20}_{-53.87}$	&	$131.25^{+66.55}_{-79.03}$	\\
$\log[f_a]$	&	$[-50,0]$	&	$-44.39^{+0.05}_{-0.05}$	&	$-44.17^{+0.04}_{-0.05}$	\\
$\log[f_h]$	&	$[13,15]$	&	$14.70^{+0.01}_{-0.01}$	&	$14.70^{+0.01}_{-0.01}$	\\
$\log[f_k]$	&	$[-20,0]$	&	$-15.25^{+0.03}_{-0.04}$	&	$-15.15^{+0.03}_{-0.03}$	\\
\hline							
$\chi^2_{\rm lc}/\rm{dof}$	&	-	&	$12761/280$	&	$17368/292$	\\
$\chi^2_{\rm vel}/\rm{dof}$	&	-	&	$19/7$	&	$36/7$	\\
$\chi^2_{\rm tot}/\rm{dof}$	&	-	&	$12780/287$	&	$17404/299$	\\
\enddata
\tablecomments{The uncertainties of the best-fitting parameters are measured as $1\sigma$ confidence ranges.}
\end{deluxetable}

\clearpage

\section{Corner plot} \label{sec:corner}

\begin{figure}[b]
\centering
\includegraphics[width=1.0\textwidth,angle=0]{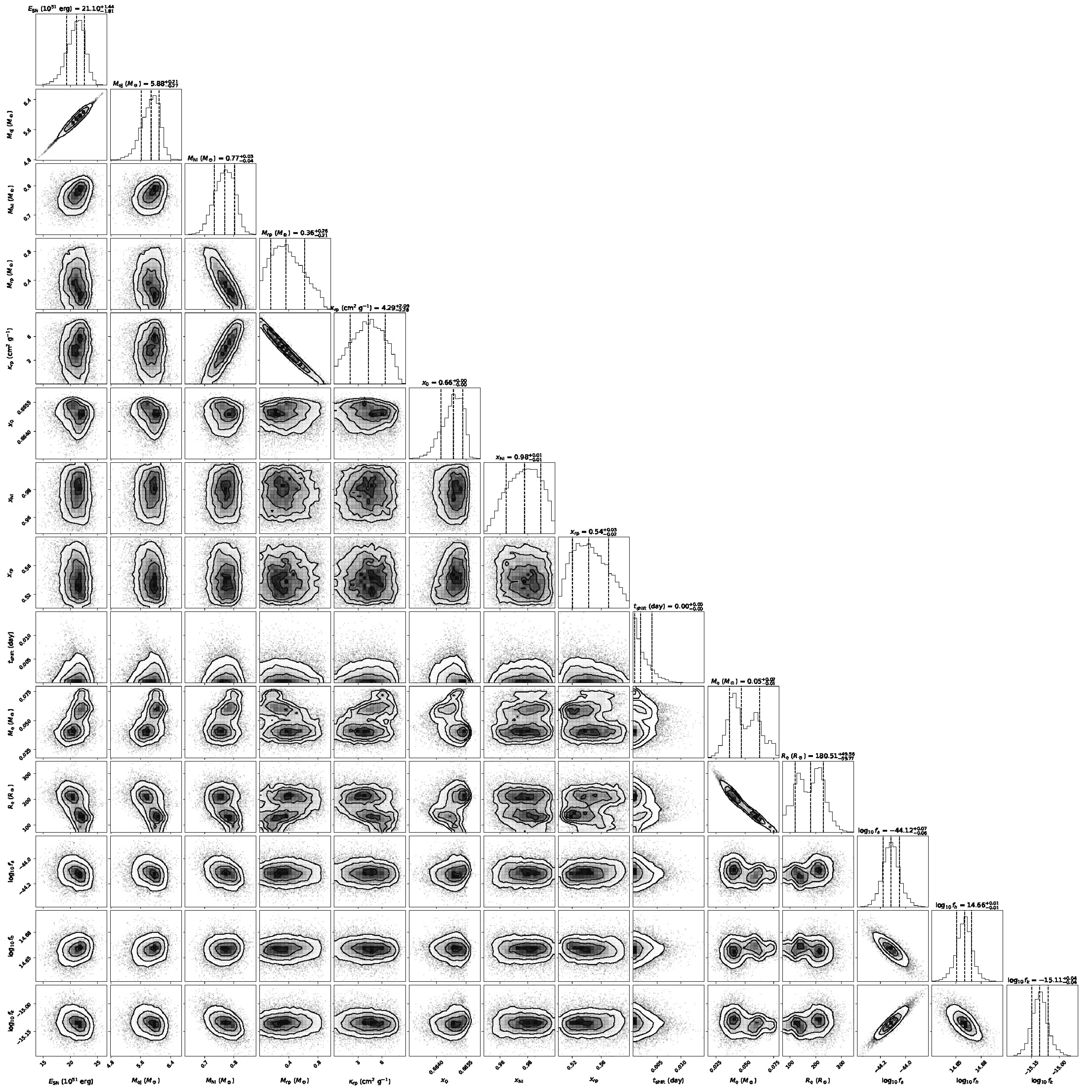}
\caption{Corner plots showing the one and two dimensional projections of the posterior probability distributions of parameters for the $r$-process-enriched model.
\label{fig:cornerA}}
\end{figure}

\begin{figure}[b]
\centering
\includegraphics[width=1.0\textwidth,angle=0]{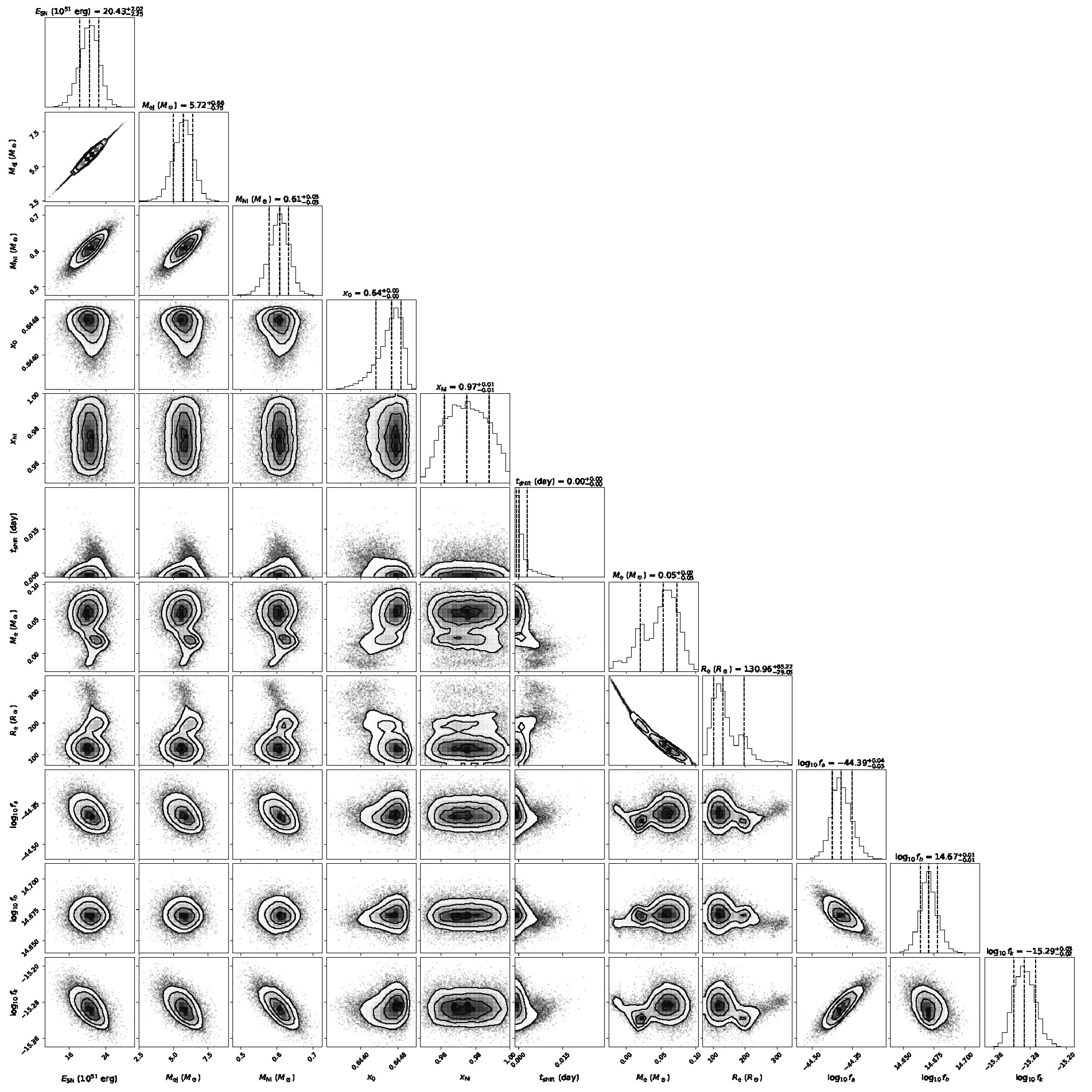}
\caption{Corner plots showing the one and two dimensional projections of the posterior probability distributions of parameters for the $r$-process-free model.
\label{fig:cornerB}}
\end{figure}

We show the corner plots for the $r$-process-enriched and  $r$-process-free models in Figure \ref{fig:cornerA} and \ref{fig:cornerB}, respectively.

\end{document}